\documentclass[a4paper,11pt]{article}
\usepackage{physics}
\usepackage[dvips]{graphicx}
\usepackage{amsmath}
\usepackage{amssymb}
\usepackage{xcolor}
\usepackage{comment}

\makeatletter
 \renewcommand{\theequation}{%
 \thesection.\arabic{equation}}
 \@addtoreset{equation}{section}
\makeatother

\def\nn{\nonumber}

\def\l{\left}
\def\r{\right}

\newcommand{\VEV}[1]{\langle{#1}\rangle}
\newcommand{\sla}[1]{\ooalign{\hfil/\hfil\crcr$#1$}} 

\begin{document}

\begin{titlepage}

\begin{flushright}
KYUSHU-HET-284 
\end{flushright}

\centering

{\Large\textbf{
Nambu-Goldstone Modes in Magnetized $T^{2n}$ Extra Dimensions
}}
\vspace{1cm}

\renewcommand{\thefootnote}{\fnsymbol{footnote}}
Takuya Hirose${}^{1}$\footnote[1]{hirose.takuya@phys.kyushu-u.ac.jp},
Hajime Otsuka${}^{1}$\footnote[2]{otsuka.hajime@phys.kyushu-u.ac.jp},
Koji Tsumura${}^{1}$\footnote[3]{tsumura.koji@phys.kyushu-u.ac.jp},
and Yoshiki Uchida${}^{2,3}$\footnote[4]{uchida.yoshiki@m.scnu.edu.cn}
\vspace{5mm}

\textit{
 $^1${Department of Physics, Kyushu University,\\
 744 Motooka, Nishi-ku, Fukuoka, 819-0395, Japan}\\
 $^2${Key Laboratory of Atomic and Subatomic Structure and Quantum Control (MOE),\\
 Guangdong Basic Research Center of Excellence for Structure and Fundamental Interactions of Matter, Institute of Quantum Matter,\\ 
 South China Normal University, Guangzhou 510006, China}\\
 $^3${Guangdong-Hong Kong Joint Laboratory of Quantum Matter, Guangdong Provincial Key Laboratory of Nuclear Science, Southern Nuclear Science Computing Center,\\ 
 South China Normal University, Guangzhou 510006, China}
}

\date{\today}

\abstract{
We consider a $U(1)$ gauge theory on $M^4\times T^4$ with background magnetic fluxes. 
We show that a theory including arbitrary fluxes can always be studied in a theory involving 
only diagonal fluxes by appropriate coordinate transformations. 
It is found that the number of independent magnetic fluxes is equal to the rank of the classical value 
of the field strength matrix, ${\rm rank}\VEV{F}$. 
The number of massless zero modes induced from extra components of higher-dimensional gauge field 
(Wilson-line scalar field), is also determined by ${\rm rank}\VEV{F}$. 
We explicitly confirm that the quantum corrections due to the matter fermion to the squared mass 
of Wilson-line scalar field cancel out at one-loop level. 
For this purpose, we derive the fermion mass spectrum on $M^4\times T^4$ with arbitrary fluxes. 
By taking the flux diagonal basis, creation and annihilation operators for Kaluza-Klein quantum numbers 
are defined appropriately.
Our results are easily generalized to the case of $M^4\times T^{2n}~(n\geq3)$. 
}

\end{titlepage}

\section{Introduction}
The Standard Model of particle physics (SM) well describes the behavior of 
elementary particles up to TeV 
scale. 
However, there are still puzzles to be answered. 
One of them is the hierarchy problem.
This problem suggests that the electroweak scale is separated from a scale of 
an ultraviolet (UV) theory, e.g. the Kaluza-Klein (KK) scale and the Planck scale.
Since quantum corrections to the mass of the SM Higgs boson are sensitive to 
the UV cutoff scale, the measured mass of the Higgs boson cannot be realized 
without unnatural fine-tuning among parameters.
To solve the hierarchy problem without the fine-tuning, we must clarify 
a UV model as a model of beyond the SM (BSM).

As a solution to the hierarchy problem, the higher-dimensional gauge theories have been studied \cite{Hosotani:1983xw,Hatanaka:1998yp}. 
In these theories, zero modes induced from extra components of 
higher-dimensional gauge field (so-called Wilson-line (WL) scalar field) 
by the compactification of the extra dimensions can be identified 
as a Higgs boson in four dimensions (4D).
Regarding a WL scalar field as a Higgs boson, it obtains the finite mass by the quantum corrections \cite{Hatanaka:1998yp,Antoniadis:2001cv,Hosotani:2007kn,Alfaro:2006is}. 
Since the finite Higgs boson mass is scaled by an inverse of 
the size of extra-dimensional space $L^{-1}$, the finite Higgs mass grows as $L^{-1}$ increases. 
If there are no signs of new physics BSM up to a large UV scale, the scale again causes a fine-tuning 
of parameters resulting in the hierarchy problem even in higher-dimensional gauge theories.

A higher-dimensional theory in a magnetic flux compactification (see for an early work \cite{Witten:1984dg}) may give us a solution to the hierarchy problem 
even if $L^{-1}$ is large\footnote{In addition, flux compactifications can give the origin of the flavor structure, the chiral structure of fermions, the number of generations of fermions, and the hierarchy of Yukawa couplings of quarks and leptons, where the SM does not 
explain these problems from the first principle.  Indeed, concrete examples were given in string theory, e.g., toroidal compactifications of type I and type IIB superstring theories with magnetized D-branes \cite{Blumenhagen:2000wh,Angelantonj:2000hi,Blumenhagen:2000ea,
Cremades:2004wa} and $E_8\times E_8'$ heterotic string theory on smooth Calabi-Yau threefolds with line bundles \cite{Blumenhagen:2005ga,Anderson:2012yf} as well as $SO(32)$ heterotic string theory \cite{Blumenhagen:2005pm,Abe:2015mua,Otsuka:2018oyf}. For more details, see, e.g., Refs. \cite{Blumenhagen:2006ci,Ibanez:2012zz}.}. 
It is known that the introduction of a background magnetic flux affects the WL scalar field 
in 4D effective theories. 
A remarkable feature of the WL scalar field in the flux compactification is 
the cancellation of quantum corrections to the mass of the WL scalar field. 
This has been shown in Abelian gauge theory on $M^4\times T^2$ at the one-loop level \cite{Buchmuller:2016gib,Buchmuller:2018eog,Ghilencea:2017jmh} and at a higher-loop level \cite{Honda:2019ema}, and in non-Abelian gauge theory on $M^4\times T^2$ at the one-loop level \cite{Hirose:2019ywp}. 
The cancellation of quantum corrections to the mass of the WL scalar field is associated with 
a shift symmetry.
This shift symmetry is understood as a remnant of translational symmetry of the coordinates 
on the torus after introducing the magnetic flux. 
Thus, the WL scalar field may be identified as a Higgs boson without conflicting 
the hierarchy problem. 
On the other hand, the WL scalar becomes an exact massless Nambu-Goldstone (NG) mode 
in the flux compactification, and further extensions are necessary in order to have a finite Higgs boson mass. 
It has been shown that the quantum corrections to the mass of the WL scalar field can be finite by introducing a certain interaction term in six-dimensional scalar quantum electrodynamics \cite{Hirose:2021rit}.

When considering the higher dimensional space $T^{2n}~(n\geq2)$ as extra dimensions~\cite{Antoniadis:2009bg,Kikuchi:2020nxn,Hoshiya:2020hki,Kikuchi:2022lfv,Kikuchi:2022psj,Kikuchi:2023awm,Kikuchi:2023awe}, the impact of magnetic fluxes on the 4D theory has not yet been clarified. 
In such an extra-dimensional space, not only can multiple magnetic fluxes be introduced into 
the diagonal component of the gauge field strength, but also into the off-diagonal component. 
The impact of these different types of magnetic flux on the 4D theory must be clarified; how many massless NG modes appear in the 4D theory will greatly affect the physics of the 4D theory. 

This paper is organized as follows: in Sec.~\ref{sec:t4}, we review the magnetized $T^4$ model. 
In Sec.~\ref{sec:ngb}, we discuss the number of NG modes based on the model with arbitrary magnetic fluxes. 
In Sec.~\ref{sec:DiracEq}, we solve the Dirac equation on $T^4$ in the flux diagonal basis 
to derive the mass spectrum of the fermions. 
The relevant interactions to the one-loop calculation are given in \ref{app-A}. 
Our developed method is extended to $T^{2n}~(n\geq3)$ in \ref{app-B}. 
In Sec.~\ref{sec:1loopMass}, we calculate the quantum corrections to 
the mass of the WL scalar field in $T^4$ theory, and verify its cancellation. 
We summarize our study in Sec.~\ref{sec:summary}. 
%

\section{Magnetized $T^4$ model}
\label{sec:t4}
Following \cite{Cremades:2004wa,Antoniadis:2009bg,Kikuchi:2022lfv,Kikuchi:2022psj},
we introduce four-dimensional torus $T^4\simeq \mathbb{C}^2/\Lambda$,
where $\Lambda$ is a lattice spanned by four independent lattice vectors 
$\vec{e}_{I}^{}$, $(I=1,2,3,4)$ in $\mathbb{C}^2$. 
In this paper, we are interested in cases when at least two of the vectors are perpendicular. 
By a suitable orthogonal rotation, we can write $\vec{e}_{I}^{}$'s of the form 
\begin{align}
\label{eq:t4_lattice-basis}
&\vec{e}_1 
= 
2\pi R_1
\l(  \begin{array}{c}
     1
     \\
     0
     \end{array}
\r) \, ,
\qquad
\vec{e}_2 
= 
2\pi 
\l(  \begin{array}{c}
     R_1\, \Omega^1{}_{1}
     \\
     R_2\, \Omega^2{}_{1}
     \end{array}
\r) \, ,  
\nn\\    
&\vec{e}_3
= 
2\pi R_2
\l(  \begin{array}{c}
     0
     \\
     1
     \end{array}
\r) \, ,   
\qquad
\vec{e}_4 
= 
2\pi
\l( \begin{array}{c}
    R_1\, \Omega^1{}_{2}
    \\
    R_2\, \Omega^2{}_{2} 
    \end{array}
\r) \, ,
\end{align}
where the complex structure $\Omega^i{}_{j}\in \mathbb{C}^2$ $(i,j=1,2)$ and scale factors $R_1$, $R_2> 0$.
We introduce $x^i$ and $y^i$ as real and dimensionless coordinates along the lattice vectors of the torus. 
We introduce dimensionless complex coordinates $z^i,~\bar{z}^{\,\bar{i}}$ $(\bar{i}=1,2)$ as
\begin{align}
\label{eq:flux_xyz-relation}
z^i 
= x^i + \Omega^i{}_j\, y^j \, ,
\quad
\bar{z}^{\,\bar{i}} 
= x^{\,\bar{i}} 
+ \overline{\Omega}^{\,\bar{i}}{}_{\bar{j}} \, y^{\,\bar{j}} \, ,
\end{align}
where $\overline{\Omega}^{\,\bar{i}}{}_{\bar{j}}$ is a complex conjugate of $\Omega^i{}_j$.
The square of the line element is given by
\begin{align}
{\rm d}s^2 
= 2 h_{i\bar{j}} {\rm d}z^i {\rm d}\bar{z}^{\bar{j}} \, ,
\end{align}
with metric
\begin{align}
h_{i\bar{j}} = \frac{1}{2}(2\pi R_i)^2 \delta_{i\bar{j}} \, ,
\end{align}
where summation is not taken for $i$.
Gamma matrices on the complex coordinates of $T^4$ are given by
\begin{align}
&(2\pi R_1)\Gamma^{z^1}
= \sigma_+ \otimes \sigma^3
= 
  \l(  \begin{array}{cccc}
        0 & 2 & {} & {}
        \\
        0 & 0 & {} & {}
        \\
        {} & {} & 0 & -2
        \\
        {} & {} & 0 & 0
        \end{array}
   \r) \, ,
\qquad
(2\pi R_2)
\Gamma^{z^2}
= I \otimes \sigma_+
=
  \l(  \begin{array}{cccc}
        {} & {} & 2 & 0
        \\
        {} & {} & 0 & 2
        \\
        0 & 0 & {} & {}
        \\
        0 & 0 & {} & {}
        \end{array}
   \r) \, ,
\nn\\
&(2\pi R_1)\Gamma^{\bar{z}^1}
= \sigma_- \otimes \sigma^3
=
  \l(  \begin{array}{cccc}
        0 & 0 & {} & {}
        \\
        2 & 0 & {} & {}
        \\
        {} & {} & 0 & 0
        \\
        {} & {} & -2 & 0
        \end{array}
   \r) \, ,
\qquad
(2\pi R_2)
\Gamma^{\bar{z}^2}
= I \otimes \sigma_-
= 
  \l(  \begin{array}{cccc}
        {} & {} & 0 & 0
        \\
        {} & {} & 0 & 0
        \\
        2 & 0 & {} & {}
        \\
        0 & 2 & {} & {}
        \end{array}
   \r) \, , 
\end{align}
satisfying $\{\Gamma^{z^i}, \Gamma^{\bar{z}^{\bar{j}}}\} = 2h^{i\bar{j}}$ where $h^{i\bar{j}}$ is the inverse of $h_{i\bar{j}}$.
Here, $\sigma^a~(a=1,2,3)$ denote the Pauli matrices, $\sigma_\pm = \sigma^1 \pm i\sigma^2$ 
and $I$ is a $2\times 2$ identity matrix.
The chirality matrix $\Gamma^5$ is given by
\begin{align}
\label{eq:flux_Gamma5}
\Gamma^5
= \sigma^3 \otimes \sigma^3 
= \l(  \begin{array}{cccc}
        1 & {} & {} & {}
        \\
        {} & -1 & {} & {}
        \\
        {} & {} & -1 & {}
        \\
        {} & {} & {} & 1
        \end{array}
   \r) \, .      
\end{align}
\renewcommand{\arraystretch}{1.0}

We consider a $U(1)$ gauge theory. The field strength on $T^4$ is given by
\begin{align}
F
=
\frac{1}{2} F_{x^ix^j} {\rm d}x^i \wedge {\rm d}x^j
+
\frac{1}{2} F_{y^iy^j} {\rm d}y^i \wedge {\rm d}y^j
+
F_{x^iy^j} {\rm d}x^i \wedge {\rm d}y^j \, ,
\end{align}
where each element is expressed in terms of gauge fields $A_{x^i}$ and $A_{y^i}$ as
\begin{align}
    F_{x^ix^j} 
    = \partial_{x^i}A_{x^j} - \partial_{x^j}A_{x^i} \, ,
    \quad
    F_{y^iy^j} 
    = \partial_{y^i}A_{y^j} - \partial_{y^j}A_{y^i} \, ,
    \quad
    F_{x^iy^j} 
    = \partial_{x^i}A_{y^j} - \partial_{y^j}A_{x^i} \, .
\end{align}
In the following, we introduce background magnetic flux.
The background magnetic flux on $T^4$ is expressed as the classical value of a field strength as follows,
\begin{align}
\label{eq:flux_F-real}
\VEV{F}
=
\frac{1}{2} p^{}_{x^ix^j} {\rm d}x^i \wedge {\rm d}x^j
+
\frac{1}{2} p^{}_{y^iy^j} {\rm d}y^i \wedge {\rm d}y^j
+
p^{}_{x^iy^j} {\rm d}x^i \wedge {\rm d}y^j \, ,
\end{align}
with
\begin{align}
p_{x^ix^j} = \VEV{F_{x^ix^j}}  \, ,
\quad
p_{y^iy^j} = \VEV{F_{y^iy^j}}  \, ,
\quad
p_{x^iy^j} = \VEV{F_{x^iy^j}}  \, ,
\end{align}
where $p^{}_{x^ix^j}$, $p^{}_{y^iy^j}$, and $p^{}_{x^iy^j}$ are real parameters determined later.
Rewriting the classical value of the field strength \eqref{eq:flux_F-real} in terms of complex coordinate 
$z^i$ and $\bar{z}^{\bar{i}}$, we get
\begin{align}
\label{eq:flux_F-classical-1}
\VEV{F} 
= 
\frac{1}{2} \VEV{F_{z^i z^j}} \,{\rm d}z^i \wedge {\rm d}z^j
+
\frac{1}{2} \VEV{F_{\bar{z}^{\,\bar{i}} \bar{z}^{\,\bar{j}}}} \,{\rm d}\bar{z}^{\,\bar{i}} \wedge {\rm d}\bar{z}^{\,\bar{j}}
+
\VEV{F_{z^i\bar{z}^{\bar{j}}}} (i{\rm d}z^i \wedge \,{\rm d}\bar{z}^{\,\bar{j}}) \, ,
\end{align}
with
\begin{align}
\label{eq:flux_Fzz}
&\VEV{F_{z^iz^j}}
=
{(\bar{\Omega} - \Omega)^{-1}}^T
\big( 
\bar{\Omega}^T p^{}_{xx} \bar{\Omega} 
+p^{}_{yy}
+p_{xy}^T \bar{\Omega}
-\bar{\Omega}^T p^{}_{xy}
\big)
(\bar{\Omega} - \Omega)^{-1} \, ,
\\
\label{eq:flux_Fzbarzbar}
&\VEV{F_{\bar{z}^{\bar{i}}\bar{z}^{\bar{j}}}}
=
{(\bar{\Omega} - \Omega)^{-1}}^T
\big(
\Omega^T p^{}_{xx} \Omega
+p^{}_{yy}
+p_{xy}^T \Omega
-\Omega^T p^{}_{xy}
\big)
(\bar{\Omega} - \Omega)^{-1} \, ,
\\
&\VEV{F_{z^i\bar{z}^{\bar{j}}}}
=
i{(\bar{\Omega} - \Omega)^{-1}}^T
\big(
\bar{\Omega}^T p^{}_{xx} \Omega
+p^{}_{yy}
+p_{xy}^T \Omega
-\bar{\Omega}^Tp^{}_{xy}
\big)
(\bar{\Omega} - \Omega)^{-1} \, .
\end{align}
We note that $\VEV{F_{z^i z^j}}$ and $\VEV{F_{\bar{z}^{\bar{i}}\bar{z}^{\bar{j}}}}$ vanish 
if supersymmetry is imposed \cite{Candelas:1985en}. 
In this paper, we do not assume supersymmetry. 
However, since the value of the background flux can be freely chosen as long as it satisfies the equations of motion, 
we assume 
\begin{align}
\label{eq:flux_susy-condition}
\Omega^T p^{}_{xx}\, \Omega
+p^{}_{yy}
+p_{xy}^T\Omega
-\Omega^T p^{}_{xy}
=0 \, .
\end{align}
In this case, $\VEV{F}$ is simplified to
\begin{align}
\label{eq:flux_F-classical-2}
\VEV{F}
=
-\frac{1}{2}
[(p^{}_{xx}\Omega - p^{}_{xy}) 
({\rm Im} \Omega)^{-1}]_{i\bar{j}}
(i{\rm d}z^i \wedge {\rm d}\bar{z}^{\bar{j}}) \, .
\end{align}
To further simplify the discussion, we choose $p^{}_{xx}$ and $p^{}_{yy}$ to be zero values;
\begin{align}
\label{eq:flux_pxx-pyy-vanish}
p^{}_{xx} = p^{}_{yy} = 0 \, .
\end{align}
Then, $\VEV{F}$ becomes
\begin{align}
\label{eq:flux_F-classical-3}
\VEV{F} 
= 
\frac{1}{2} [ p^{}_{xy} ({\rm Im} \Omega)^{-1} ]_{i\bar{j}} (i{\rm d}z^i \wedge {\rm d}\bar{z}^{\bar{j}}) \, .
\end{align}
Note that, by imposing Eq.~\eqref{eq:flux_pxx-pyy-vanish},
the condition \eqref{eq:flux_susy-condition} becomes
\begin{align}
\label{eq:flux_susy-condition-2}
(p_{xy}^T \Omega)^T
=
p_{xy}^T \Omega \, .
\end{align}
The value of the background flux is quantized by the boundary conditions as \cite{Antoniadis:2009bg}
\begin{align}
\label{eq:flux_pxy-quantized}
p^{}_{xy} = 2\pi N^T 
\end{align}
with $N$ being a $2\times 2$ integer matrix.
By substituting Eq.~\eqref{eq:flux_pxy-quantized} into Eq.~\eqref{eq:flux_F-classical-3}, we get
\begin{align}
\label{eq:flux_Fzzbar-classical}
\VEV{F_{z^i \bar{z}^{\bar{j}}}}
=
\pi
[\, N^T({\rm Im} \Omega)^{-1}]^{}_{i\bar{j}} \, .
\end{align}
From Eq.~\eqref{eq:flux_Fzzbar-classical}, we see that the classical value of $F_{z^i\bar{z}^{\bar{j}}}$ 
takes a constant value. 
Hence, it is easy to see that $\VEV{F}$ satisfies the equation of motion. 
Note that, by applying Eq.~\eqref{eq:flux_pxy-quantized}, the condition \eqref{eq:flux_susy-condition-2} becomes
\begin{align}
\label{eq:flux_susy-condition-3}
(N\Omega)^T
=
N\Omega \, .
\end{align}
The field strength can be expressed using gauge fields as 
\begin{align}
F_{z^i\bar{z}^{\bar{j}}}
=\frac{1}{i}(\partial_{z^i} A_{\bar{z}^{\bar{j}}} - \partial_{\bar{z}^{\bar{j}}} A_{z^i}) \, .
\end{align}
It is easy to check that Eq.~\eqref{eq:flux_Fzzbar-classical} can be reproduced 
by taking the classical value of the gauge component as follows,
\begin{align}
\label{eq:flux_Az-Azbar-classical}
&\VEV{A_{z^i}}
=
-i\frac{\pi}{2} \bar{z}^{\bar{j}} [\,N^T ({\rm Im} \Omega)^{-1}]^{}_{\bar{j}i} \, , 
\quad
\VEV{A_{\bar{z}^{\bar{i}}}}
=
i\frac{\pi}{2} z^j [\,N^T ({\rm Im} \Omega)^{-1}]^{}_{j\bar{i}} \, .
\end{align}
Giving Eq.~\eqref{eq:flux_Az-Azbar-classical}, $A_{z^i}$ and $A_{\bar{z}^{\bar{i}}}$ 
can be expanded around the flux backgrounds as 
\begin{align}
\label{WLscalar}
A_{z^i}
=\VEV{A_{z^i}}+\frac{\varphi_{z^i}}{\sqrt{2}}\,,\quad
A_{\bar{z}^{\bar{i}}}
=\VEV{A_{\bar{z}^{\bar{i}}}}+\frac{\varphi_{\bar{z}^{\bar{i}}}}{\sqrt{2}}\,.
\end{align}
The factor of $1/\sqrt{2}$ is introduced to canonically normalize $\varphi_{z^i}$ and $\varphi_{\bar{z}^{\bar{i}}}$. 
Hereafter, we call $\varphi_{z^i}~(\varphi_{\bar{z}^{\bar{i}}})$ WL scalar fields. 
It is also clear that because $\VEV{F_{z^i\bar{z}^{\bar{j}}}}$ is real, the following relation holds,
\begin{align}
\label{eq:flux_F*}
\VEV{F_{z^i\bar{z}^{\bar{j}}}}
=\overline{\VEV{F_{z^i\bar{z}^{\bar{j}}}}}
=-\VEV{F_{\bar{z}^{\bar{i}}z^j}}
=\VEV{F_{z^j\bar{z}^{\bar{i}}}} \, . 
\end{align}
That is, $\VEV{F_{z^i\bar{z}^{\bar{j}}}}$ is symmetric. 

Before closing this section, we introduce the eight-dimensional (8D) Weyl fermion $\Psi$.
Free lagrangian for $\Psi$ with background magnetic fluxes is given by 
\begin{align}
\label{eq:DiracEq_lag8D}
\mathcal{L}^{\rm 8D}_{\rm free}
=&\,
\overline{\Psi} i\Gamma^\mu \partial_\mu \Psi
+
\overline{\Psi} 
i \sla{\mathcal{D}}_z \Psi \, ,
\end{align}
with
\begin{align}
\label{eq:DiracEq_iD}
i\sla{\mathcal{D}}_z
&=
i\l(  
\Gamma^{z^j} \mathcal{D}_{z^j} 
+ \Gamma^{\bar{z}^{\bar{j}}} \mathcal{D}_{\bar{z}^{\bar{j}}}  
\r)
=
2i
\l(  \begin{array}{cccc}
     0 & \rho_1\mathcal{D}_{z^1} & \rho_2\mathcal{D}_{z^2} & 0
     \\
     \rho_1\mathcal{D}_{\bar{z}^1} & 0 & 0 & \rho_2\mathcal{D}_{z^2}
     \\
     \rho_2\mathcal{D}_{\bar{z}^2} & 0 & 0 & -\rho_1\mathcal{D}_{z^1}
     \\
     0 & \rho_2\mathcal{D}_{\bar{z}^2} & -\rho_1\mathcal{D}_{\bar{z}^1} & 0
     \end{array}
\r)  \, ,
\end{align}
where $\rho_j=1/(2\pi R_j)$. 
The covariant derivatives with the gauge field replaced by its classical value are given by
\begin{align}
\label{eq:DiracEq_covader}
\mathcal{D}_{z^i}\Psi
=
\Big( \partial_{z^i} + iq \VEV{A_{z^i}} \Big) \Psi \, ,
\quad
\mathcal{D}_{\bar{z}^{\bar{i}}}\Psi
=
\Big( \partial_{\bar{z}^{\bar{i}}} + iq \VEV{A_{\bar{z}^{\bar{i}}}} \Big) \Psi \, ,
\end{align}
where $q$ is the 8D gauge coupling times a charge of $\Psi$. 
The $\Psi$ is expressed in terms of four-component  fields 
$\big(\psi^{\bf 1}_L, \psi^{\bf 2}_R, \psi^{\bf 3}_R, \psi^{\bf 4}_L\big)$ as 
\begin{align}
\label{eq:DiracEq_Psi8D}
\Psi
=
\l(  \begin{array}{c}
     \psi^{\bf 1}_L
     \\
     \psi^{\bf 2}_R
     \\
     \psi^{\bf 3}_R
     \\
     \psi^{\bf 4}_L
     \end{array}
\r) \, ,
\end{align}
where the subscripts $L$ and $R$, respectively, correspond to the left- and right-handed 
chiralities in 4D after integrating the extra dimension $T^4$.

\section{Counting Nambu-Goldstone Modes}
\label{sec:ngb}
%
As pointed out in Refs.~\cite{Buchmuller:2018eog,Honda:2019ema}, the WL scalar fields become NG modes 
since the introduction of the magnetic flux leaves a constant shift symmetry for the WL scalar field. 
This shift symmetry is a remnant of the spontaneous breaking of the translation of the coordinates 
on the torus due to the magnetic flux.
It is confirmed that the quantum corrections to the mass of the WL scalar field are canceled out.

In the case of the higher dimensions of $T^{2n}~(n\geq 2)$,
contrary to $T^2$, multiple magnetic fluxes can be introduced to the theory. 
In general, the field strength matrix $\VEV{F}$ also has non-zero values for the off-diagonal elements. 
In the following, we extend the formulation of the shift symmetry on $T^{2}$ in  Refs.~\cite{Buchmuller:2018eog,Honda:2019ema} to those in $T^{4}$.

The original 8D lagrangian in a $U(1)$ gauge theory is invariant under the following transformations, 
which are obtained by combining gauge and translational transformations on the $T^{4}$ with magnetic fluxes. 
The translational operators in $T^{4}$ are expressed as 
$\delta_{T^{4}}=
\epsilon_j\partial_{z^j}+\bar{\epsilon}_j\partial_{\bar{z}^{\bar{j}}}$, 
where $\epsilon_j,~\bar{\epsilon}_j$ are infinitesimal parameters. 
First, we consider the infinitesimal transformation for the WL scalar field. 
The breaking of translational symmetry by the magnetic flux can be compensated by a constant shift. 
Thus, the translational transformation for the WL scalar field is described as
\begin{align}
\label{translation:WLscalar}
\delta_{T^{2n}}\varphi_{z^i}=(\epsilon_j\partial_{z^j}+\bar{\epsilon}_j\partial_{\bar{z}^{\bar{j}}})\varphi_{z^i}-\frac{i}{\sqrt{2}}\bar{\epsilon}_j\VEV{F_{z^j\bar{z}^{\bar{i}}}}\,.
\end{align}
The gauge transformations for the WL scalar field are given by
\begin{align}
\label{gauge:WLscalar}
\delta_{\Lambda}\varphi_{z^i}=-\sqrt{2}\partial_{z^i}\Lambda=-\frac{i}{\sqrt{2}}\bar{\alpha}_j\VEV{F_{z^j\bar{z}^{\bar{i}}}},\quad
\Lambda=\frac{i}{2}\VEV{F_{z^j\bar{z}^{\bar{i}}}}(\bar{\alpha}_j^{} z^{i}-\alpha_j^{} \bar{z}^{\bar{i}})\,, 
\end{align}
where $\alpha_j^{}$, $\bar{\alpha}_j^{}$ are complex parameters. 
Second, we consider the infinitesimal transformation for the fermion.
The translation of the fermion is given by
\begin{align}
\label{tranlation:fermion}
\delta_{T^{2n}}\Psi=(\epsilon_j\partial_{z^j}+\bar{\epsilon}_j\partial_{\bar{z}^{\bar{j}}})\Psi\,.
\end{align}
The gauge transformation for the fermion is given by 
\begin{align}
\label{gauge:fermion}
\delta_\Lambda\Psi=iq\Lambda\Psi=iq\big(\alpha_j \VEV{A_{z^j}}+\bar{\alpha}_j \VEV{A_{\bar{z}^{\bar{j}}}}\big)\Psi\,. 
\end{align}
From Eq.\eqref{tranlation:fermion} and Eq.\eqref{gauge:fermion} with $\epsilon_j=\alpha_j$, 
a combined transformation for the fermion can be expressed by using covariant derivatives: 
\begin{align}
\delta\Psi=(\delta_{T^{2n}}+\delta_\Lambda)\Psi=\big( \epsilon_j \mathcal{D}_{z^j} + \bar{\epsilon}_j \mathcal{D}_{\bar{z}^{\bar{j}}}  \big) \Psi\,.
\end{align}
On the other hand, from Eq.\eqref{translation:WLscalar} and Eq.\eqref{gauge:WLscalar} with $\epsilon_j=\alpha_j$, the combined transformation for the WL scalar field corresponds to a constant shift:
\begin{align}
\label{total:WLscalar}
\delta\varphi_{z^i}=-\sqrt{2}i\bar{\epsilon}_j\VEV{F_{z^j\bar{z}^{\bar{i}}}}\,.
\end{align}
Note that derivative terms in Eq.\eqref{translation:WLscalar} vanish since $\varphi_{z^i}$ are 
independent of complex coordinates. 
Thus, the WL scalar fields are identified as NG modes. 
Conversely, if the WL scalar fields do not feel the magnetic flux in Eq.\eqref{total:WLscalar}, 
they have a mass and cannot be an NG mode.\footnote{In the case of $T^2$ without a flux, the WL scalar field receives the quantum correction to the mass.
See \cite{Antoniadis:2001cv,Alfaro:2006is,Buchmuller:2016gib}.}
This formulation in $T^4$ is easily extended to one in $T^{2n}~(n\ge2)$, 
e.g., the index $i$ of complex coordinates runs from one to three in $T^6$.

Since the constant shift in Eq.\eqref{total:WLscalar} associates with the magnetic flux,
it is natural to expect that the number of massless WL scalar fields is controlled 
by the number of independent magnetic fluxes. 
To extract the information of the independent magnetic fluxes, we consider the diagonalization 
of the flux matrix $\VEV{F}$. 
The field strength two-form $F$ is invariant under the coordinate transformation as 
\begin{align}
F
= F_{z^i\bar{z}^{\,\bar{j}}} {\rm d}z^i \wedge {\rm d}\bar{z}^{\,\bar{j}}
= (U\widetilde{F}U^T)_{i\bar{j}} {\rm d}z^i \wedge {\rm d}\bar{z}^{\,\bar{j}}
= \widetilde{F}_{z^i\bar{z}^{\,\bar{j}}} {\rm d}\tilde{z}^i \wedge {\rm d}\bar{\tilde{z}}^{\,\bar{j}}, 
\end{align}
where the coordinate transformations of $z$ as well as $\Omega$ are given by
\begin{align}
\vec{\tilde{z}}
= U^T \vec{z}
= U^T \vec{x} + U^T \Omega U U^T \vec{y}
= \vec{\tilde{x}} + \widetilde{\Omega} \vec{\tilde{y}} \, ,
\end{align}
with 
\begin{align}
\label{eq:DiracEq_UOmegaU}
\widetilde{\Omega} = U^T \Omega U \, ,
\qquad
\vec{\tilde{x}} = U^T \vec{x} \, ,
\qquad
\vec{\tilde{y}} = U^T \vec{y} \, .
\end{align}
Using this coordinate transformation, we take the flux diagonal basis as 
\begin{align}
\label{eq:DiracEq_UFU}
\VEV{\widetilde{F}}
= U^T \VEV{F} U \, ,
\qquad
U
=
\l(  \begin{array}{cc}
     \cos\theta & \sin\theta
     \\
     -\sin\theta & \cos\theta
     \end{array}
\r) \, ,  
\end{align}
where $\VEV{\widetilde{F}}$ is a diagonal matrix, and 
\begin{align}
\label{eq:DiracEq_mixing}
\tan2\theta= 
\frac{2\VEV{F_{z^2\bar{z}^1}}}{|\VEV{F_{z^1\bar{z}^1}}-\VEV{F_{z^2\bar{z}^2}}|}\, .
\end{align}
From Eq.~\eqref{eq:flux_Fzzbar-classical}, we define $N$ in the new basis as
\begin{align}
\VEV{\widetilde{F}} = \pi \widetilde{N}^T ({\rm Im}\widetilde{\Omega})^{-1} \, ,
\end{align}
with
\begin{align}
\widetilde{N} = U^T N U \, .
\end{align}
By this procedure, any theory on $T^4$ with an arbitrary flux matrix satisfying Eq.~\eqref{eq:flux_susy-condition} 
can always be brought to a theory with only diagonal fluxes by performing the appropriate coordinate transformation.
Since the number of non-zero eigenvalues is equal to the rank of the matrix, 
the number of independent magnetic fluxes is determined by the rank of the flux matrix, ${\rm rank}\VEV{F}$. 
In the following, we will consider the theory in the flux diagonal basis. 
For the sake of simplicity, the tilde symbols, which indicate the theory in the flux diagonal basis, 
are omitted below. 
In the $M^4\times T^4$ theory, there are two complex WL scalar fields.
Eq.~\eqref{total:WLscalar} is written as
\begin{align}
\label{eq:t4_F}
\delta \varphi_{z^1}
= -\sqrt{2}i \bar{\epsilon}_1 \VEV{F_{z^1\bar{z}^1}} \, , 
\quad
\delta \varphi_{z^2}
= -\sqrt{2}i \bar{\epsilon}_2 \VEV{F_{z^2\bar{z}^2}}\, .   
\end{align}
Defining $n_\text{NG}^{}$ as the number of NG mode $\varphi_{z^i}$, we conclude
\begin{align}
\label{number=rank}
n_\text{NG}^{}={\rm rank}\VEV{F}\,.
\end{align}
There are no NG modes in a $T^4$ theory without magnetic fluxes, i.e., ${\rm rank}\VEV{F}=0$. 
In this case, the quantum corrections to the mass of two WL scalar fields both diverge.
For ${\rm rank}\VEV{F}=1$, a WL scalar field has the same properties as 
that in the previous study for the $T^2$ theory with a magnetic flux. 
Namely, the WL scalar field, which corresponds to the direction of the magnetic flux, 
becomes an NG mode. The quantum corrections to the mass of this WL scalar field cancel out. 
On the other hand, the other WL scalar field receives divergent quantum corrections to the mass. 
For a $T^4$ theory with two independent fluxes, i.e., ${\rm rank}\VEV{F}=2$, 
there are two NG modes. 
In this case, the quantum corrections to the mass of the two WL scalar fields are both canceled. 
In the next section, we will explicitly show the cancellation of the quantum corrections 
to the mass of WL scalar fields at the one-loop level. 

\section{Matter spectrum on magnetized $T^4$}
\label{sec:DiracEq}
%
In this section, we derive the mass spectrum for the fermion in 4D effective theory by solving the Dirac equation.
From Eq.~\eqref{eq:DiracEq_lag8D}, we obtain the following equation of motion,
\begin{align}
\label{eq:DiracEq_DiracEq}
(i \Gamma^\mu \partial_\mu)^2 \Psi + (i\sla{\mathcal{D}}_z)^2 \Psi
= 0 \, .
\end{align}
In the flux diagonal basis, the commutation relations for covariant derivatives are given below,
\begin{align}
\label{eq:diagonal_DDcomm-1}
&[\mathcal{D}_{z^1}, \mathcal{D}_{\bar{z}^1}] 
= -q \VEV{F_{z^1\bar{z}^1}} \, ,
\qquad
[\mathcal{D}_{z^2}, \mathcal{D}_{\bar{z}^2}] 
= -q \VEV{F_{z^2\bar{z}^2}} \, ,
\\
&
[\mathcal{D}_{z^1}, \mathcal{D}_{z^2}]
=[\mathcal{D}_{\bar{z}^1}, \mathcal{D}_{\bar{z}^2}]
=[\mathcal{D}_{z^1}, \mathcal{D}_{\bar{z}^2}] 
=[\mathcal{D}_{z^2}, \mathcal{D}_{\bar{z}^1}] 
=0 \, .
\end{align}
Based on the above commutation relations, we define creation and annihilation operators 
by normalizing the covariant derivatives,
\begin{align}
\label{eq:DiracEq_HarmonicOpe}
&a_1^{} = \frac{i}{\sqrt{q \VEV{F_{z^1\bar{z}^1}}}} \mathcal{D}_{z^1} \, ,
\qquad
a_1^\dagger = \frac{i}{\sqrt{q \VEV{F_{z^1\bar{z}^1}}}} \mathcal{D}_{\bar{z}^1} \, ,
\nn\\
&a_2^{} = \frac{i}{\sqrt{q \VEV{F_{z^2\bar{z}^2}}}} \mathcal{D}_{z^2} \, ,
\qquad
a_2^\dagger = \frac{i}{\sqrt{q \VEV{F_{z^2\bar{z}^2}}}} \mathcal{D}_{\bar{z}^2} \, ,
\end{align}
for $q\VEV{F_{z^1\bar{z}^1}}>0$ and $q\VEV{F_{z^2\bar{z}^2}}>0$. 
The above operators satisfy ordinary commutation relations $[a_i^{},a_j^\dagger]=\delta_{ij}$, and
$[a_1^{},a_2^{}]=[a_1^\dagger,a_2^\dagger]=[a_1^{},a_2^\dagger]=[a_1^\dagger,a_2]=0$.
When $q\VEV{F_{z^i\bar{z}^{\bar{i}}}}$ is negative, the definitions of the corresponding creation and annihilation operators are switched. 
Using the above annihilation operators $a_i$, the ground state mode function $\xi_{0,0}$ is defined by $a_i\, \xi_{0,0}=0$.
\footnote{
This function $\xi_{0,0}$ is described by the Riemann $\vartheta$-function 
and its degeneracy is determined by $\big|{\rm Det}\langle N\rangle\big|$\cite{Cremades:2004wa,Antoniadis:2009bg}.}
Mode functions $\xi_{n_1,n_2}$ are constructed by acting on the creation operators $a^\dag_i$ as
\begin{align}
\xi_{n_1,n_2}=\frac{1}{\sqrt{n_1! n_2!}}(a_1^\dagger)^{n_1}(a_2^\dagger)^{n_2}\xi_{0,0}\,.
\end{align}
Note that $\xi_{n_1,n_2}$ is an eigenstate for the number operator
$a_1^\dagger a_1^{}$ and $a_2^\dagger a_2^{}$,
\begin{align}
\label{eq:diraceqkk_nxi}
a_1^\dagger a_1^{} \xi_{n_1,n_2} = n_1 \xi_{n_1,n_2} \, ,
\qquad
a_2^\dagger a_2^{} \xi_{n_1,n_2} = n_2 \xi_{n_1,n_2} \, .
\end{align}
The mode function satisfies the following relations:
\begin{align}
\label{eq:DiracEq_excitation}
&a_1 \xi_{n_1,n_2} = \sqrt{n_1}\, \xi_{n_1-1,n_2} \, ,
\qquad
a_1^\dagger \xi_{n_1,n_2} = \sqrt{n_1+1}\, \xi_{n_1+1,n_2} \, ,
\nn\\
&a_2 \xi_{n_1,n_2} = \sqrt{n_2}\, \xi_{n_1,n_2-1} \, ,
\qquad
a_2^\dagger \xi_{n_1,n_2} = \sqrt{n_2+1}\, \xi_{n_1,n_2+1} \, .
\end{align}
It also satisfies an orthonormal condition,
\begin{align}
\label{eq:diagonal_orthogonal}
\int_{T^4}
{\rm d}^4z\,
\sqrt{{\rm det}h}\,
 \overline{\xi}_{n_1,n_2}\xi_{n'_1,n'_2}
=
\delta_{n_1n'_1}
\delta_{n_2n'_2} \, .
\end{align}
Note that mode functions $\xi_{n_1,n_2}$ have a normalization factor proportional to 
$\mathrm{Vol}(T^4)^{-1/2}$ with $\mathrm{Vol}(T^4)$ being the volume of $T^4$. 

In terms of creation and annihilation operators, the square of the Dirac operator is 
\footnote{The mass spectrum in $SU(n)$ gauge theory on $M^4\times T^2$ is known by Refs. \cite{Cremades:2004wa,Kojima:2023umv}.}
\begin{align}
\label{eq:DiracEq_iDsq}
(i\sla{\mathcal{D}}_z)^2
=&\, 4\,q\, 
{\rm diag}
\big( \rho_1^2 \VEV{F_{z^1\bar{z}^1}} (a_1^\dag a_1+1)+\rho_2^2\VEV{F_{z^2\bar{z}^2}} (a_2^\dag a_2+1)\,, \nn \\
        &\qquad \qquad
        \rho_1^2 \VEV{F_{z^1\bar{z}^1}} a_1^\dag a_1+\rho_2^2\VEV{F_{z^2\bar{z}^2}} (a_2^\dag a_2+1)\,, \nn \\
        &\qquad \qquad 
        \rho_1^2 \VEV{F_{z^1\bar{z}^1}} (a_1^\dag a_1+1)+\rho_2^2\VEV{F_{z^2\bar{z}^2}} a_2^\dag a_2\,, \nn \\
        &\qquad \qquad
        \rho_1^2 \VEV{F_{z^1\bar{z}^1}} a_1^\dag a_1+\rho_2^2\VEV{F_{z^2\bar{z}^2}} a_2^\dag a_2
\big)\,. 
\end{align}
Since we take $q\VEV{F_{z^1\bar{z}^1}}>0$ and $q\VEV{F_{z^2\bar{z}^2}}>0$, 
the 4D fermion $\psi_L^{\bf 4}$ is expected to have a massless mode, 
while the others are not. 
Each component of the 8D fermion \eqref{eq:DiracEq_Psi8D} is expanded by using 
the mode functions $\xi_{n_1,n_2}$ introduced above as follows.
\begin{align}
\label{eq:DiracEq_KKexpansion}
\psi^{\bf 1}_L
&=
\sum_{n_1=0}^\infty
\sum_{n_2=0}^\infty
\psi^{\bf 1}_{L,n_1+1,n_2+1}
\xi_{n_1,n_2} \, ,
\nn\\
\psi^{\bf 2}_R
&=
\sum_{n_1=0}^\infty
\sum_{n_2=0}^\infty
\psi^{\bf 2}_{R,n_1,n_2+1}
\xi_{n_1,n_2} \, ,
\nn \\
\psi^{\bf 3}_R
&=
\sum_{n_1=0}^\infty
\sum_{n_2=0}^\infty
\psi^{\bf 3}_{R,n_1+1,n_2}
\xi_{n_1,n_2} \, ,
\nn \\
\psi^{\bf 4}_L
&=
\sum_{n_1=0}^\infty
\sum_{n_2=0}^\infty
\psi^{\bf 4}_{L,n_1,n_2}
\xi_{n_1,n_2} \, ,
\end{align}
where the 4D fermions $\psi_{L/R, n_1,n_2}^{\bf 1,2,3,4}$ are defined 
to have masses of 
\begin{align}
\label{eq:DiracEq_Mn1n2}
M_{n_1,n_2}=2 \sqrt{q\Big(\rho_1^2 \VEV{F_{z^1\bar{z}^1}} n_1+\rho_2^2\VEV{F_{z^2\bar{z}^2}}n_2\Big)} \, . 
\end{align}

On the other hand, the fermion mass term in the 4D lagrangian appears in the form of a single power of $i\sla{\mathcal{D}}_z$. 
To see the fermion mass spectrum in 4D, we need to move the mass eigenbasis at the lagrangian level. 
The mass terms in the 4D effective lagrangian are given by 
\begin{align}
\mathcal{L}_{\rm mass}^{\rm 4D}
=&
\int_{T^4}
{\rm d}^4z \,\sqrt{{\rm det}h}\,
\overline{\Psi}
i\sla{\mathcal{D}}_{z} 
\Psi
\nn\\
=&\,+   
\sum_{\ell_1=1}^\infty
\sum_{n_2=0}^\infty
M_{\ell_1,0}
\Big[  \overline{\psi}^{\bf 1}_{L,\ell_1,n_2+1} 
         \psi^{\bf 2}_{R,\ell_1,n_2+1} 
         -
         \overline{\psi}^{\bf 3}_{R,\ell_1,n_2} 
         \psi^{\bf 4}_{L,\ell_1,n_2}
         \nn\\
         &\qquad\qquad\qquad\qquad\qquad
         +
         \overline{\psi}^{\bf 2}_{R,\ell_1,n_2+1} 
         \psi^{\bf 1}_{L,\ell_1,n_2+1}
         -
         \overline{\psi}^{\bf 4}_{L,\ell_1,n_2} 
         \psi^{\bf 3}_{R,\ell_1,n_2}
\Big]   
\nn\\
&\,
+
\sum_{n_1=0}^\infty
\sum_{\ell_2=1}^\infty
M_{0,\ell_2}
\Big[  \overline{\psi}^{\bf 1}_{L,n_1+1,\ell_2} 
         \psi^{\bf 3}_{R,n_1+1,\ell_2}
         +
         \overline{\psi}^{\bf 2}_{R,n_1,\ell_2} 
         \psi^{\bf 4}_{L,n_1,\ell_2}
         \nn\\
         &\qquad\qquad\qquad\qquad\qquad\quad
         +
         \overline{\psi}^{\bf 3}_{R,n_1+1,\ell_2} 
         \psi^{\bf 1}_{L,n_1+1,\ell_2}
         +
         \overline{\psi}^{\bf 4}_{L,n_1,\ell_2} 
         \psi^{\bf 2}_{R,n_1,\ell_2}
\Big] \, .
\end{align}
In performing the integration with respect to the 
extra-dimensional coordinates, we used 
Eq.~\eqref{eq:diagonal_orthogonal}.
The above lagrangian is equivalent to the description
of the following three types of Dirac fermion,
\begin{align}
\label{eq:4D_Psi12}
\psi_{\ell_1,\ell_2}^1
&=
\psi^{\bf 1}_{L,\ell_1,\ell_2}
+\psi^{\bf 2}_{R,\ell_1,\ell_2} \, ,
\nn\\
\psi_{\ell_1,\ell_2}^2
&=
\psi^{\bf 4}_{L,\ell_1,\ell_2}
+\psi^{\bf 3}_{R,\ell_1,\ell_2} \, ,
\nn\\
\chi_{\ell_1,0}
&=
\psi^{\bf 4}_{L,\ell_1,0}
+\psi^{\bf 3}_{R,\ell_1,0} \, ,
\nn \\
\chi_{0,\ell_2}
&=
-
\psi^{\bf 4}_{L,0,\ell_2}
+\psi^{\bf 2}_{R,0,\ell_2} \, ,
\end{align}
and constitute the mass terms given below,
\begin{align}
\label{eq:4D_intPsiMass-3}
\mathcal{L}_{\rm mass}^{\rm 4D}
=&\,
- \sum_{\ell_1=1}^\infty 
M_{\ell_1,0} \,\overline{\chi}_{\ell_1,0}\chi_{\ell_1,0}
-\sum_{\ell_2=1}^\infty 
M_{0,\ell_2} \, \overline{\chi}_{0,\ell_2} \chi_{0,\ell_2}
 \nn\\
 &\,
-
\sum_{\ell_1=1}^\infty
\sum_{\ell_2=1}^\infty
\l( \overline{\psi}_{\ell_1,\ell_2}^1,~ \overline{\psi}_{\ell_1,\ell_2}^2 \r)
M_\mathrm{KK}    
\l(  \begin{array}{c}
     \psi_{\ell_1,\ell_2}^1
     \\
     \psi_{\ell_1,\ell_2}^2
     \end{array}
\r) \, ,
\end{align}
where the mass matrix $M_\mathrm{KK}$ is given by
\begin{align}
\label{eq:4D_mkk}
M_\mathrm{KK}
=
\l(  \begin{array}{cc}
     -M_{\ell_1,0} & -M_{0,\ell_2}
     \\
     -M_{0,\ell_2} & M_{\ell_1,0}
     \end{array}
\r) \, .
\end{align}
The mass matrix $M_\mathrm{KK}$ is diagonalized by orthogonal matrix $U_\mathrm{KK}$:
\begin{align}
M_\mathrm{KK}^{\rm diag}
= U^{-1}_\mathrm{KK} M_\mathrm{KK}^{} U_\mathrm{KK}^{} \, ,
\end{align}
with
\begin{align}
\label{eq:1loopMass_MKKdiag}
M_\mathrm{KK}^{\rm diag}
={\rm diag}
\Big(  -M_{\ell_1,\ell_2},\, M_{\ell_1,\ell_2} \Big) \, ,\quad
U_\mathrm{KK}
=
\l(  \begin{array}{cc}
     \cos \theta_{\ell_1,\ell_2} & -\sin\theta_{\ell_1,\ell_2}
     \\
     \sin\theta_{\ell_1,\ell_2} & \cos\theta_{\ell_1,\ell_2}
     \end{array}
\r) \, .
\end{align}
The mixing angle $\theta_{\ell_1,\ell_2}$ is given by 
\begin{align}
\label{eq:4D_tan2theta}
\tan 2\theta_{\ell_1,\ell_2}
= \frac{M_{0,\ell_2}}{M_{\ell_1,0}} \, .
\end{align}
The mass eigenstates denoted by $\chi_{\ell_1,\ell_2}^1$ and $\chi_{\ell_1,\ell_2}^2$ are 
expressed by the linear combination of $\psi_{\ell_1,\ell_2}^1$ and $\psi_{\ell_1,\ell_2}^2$ 
in the interaction basis as 
\begin{align}
\label{eq:4D_Psi-basis}
\l(  \begin{array}{c}
     \gamma_5\chi_{\ell_1,\ell_2}^1
     \\
     \chi_{\ell_1,\ell_2}^2
     \end{array}
\r)
=
U_\mathrm{KK}^{-1}
\l(  \begin{array}{c}
     \psi_{\ell_1,\ell_2}^1
     \\
     \psi_{\ell_1,\ell_2}^2
     \end{array}
\r) \, .
\end{align}
We finally get
\begin{align}
\label{eq:4D_Psi-mass-eigen-2}
\mathcal{L}^{\rm 4D}_{\rm mass}
=&\,
- \sum_{\ell_1=1}^\infty 
M_{\ell_1,0} \,\overline{\chi}_{\ell_1,0}\chi_{\ell_1,0}
- \sum_{\ell_2=1}^\infty 
M_{0,\ell_2} \, \overline{\chi}_{0,\ell_2} \chi_{0,\ell_2}
\nn\\
&\,
-
\sum_{\ell_1=1}^\infty
\sum_{\ell_2=1}^\infty
M_{\ell_1,\ell_2} \, 
\Big( \overline{\chi}^1_{\ell_1,\ell_2}
\chi_{\ell_1,\ell_2}^1 
+\overline{\chi}^2_{\ell_1,\ell_2}
\chi_{\ell_1,\ell_2}^2\Big) \, .
\end{align}
From the third and fourth terms in the right hand side of Eq.~\eqref{eq:4D_Psi-mass-eigen-2}, 
we find that $\chi_{\ell_1,\ell_2}^1$ and $\chi_{\ell_1,\ell_2}^2$ are degenerate in mass. 
There is no right-handed partner for $\psi^{\bf 4}_{L,0,0}=\chi_{0,0}$.
This is consistent with the mass spectrum \eqref{eq:DiracEq_Mn1n2}
obtained from the Dirac equation.
In the same way, we also obtain Yukawa interactions between 4D WL scalar and fermion fields.
The expression for the Yukawa term is lengthy so we show it in the \ref{app-A}. 
We will use the Yukawa interactions to calculate quantum corrections to the mass of the WL scalar field in a later section.

The method shown above can be straightforwardly applied to cases with $T^{2n}$ $(n\geq 3)$. 
We show the derivation of the mass spectrum for a case with $T^6$ in \ref{app-B} as an example.

\section{One-loop corrections to the WL scalar mass}
\label{sec:1loopMass}

In Sec.~\ref{sec:ngb}, we clarified that the number of massless WL scalar fields 
is determined by ${\rm rank}\VEV{F}$.
To confirm this statement, we calculate the quantum correction to the mass of 
the WL scalar field at the one-loop level in the case of ${\rm rank}\VEV{F}=2$. 
Throughout this section, we assume that the extra-dimensional space is $T^4$. 
In the case of $T^{2n}~(n\geq3)$, the same procedure can also be applied 
using the similar fermion mass spectrum summarized in \ref{app-A}. 
We also comment on the case of ${\rm rank}\VEV{F}=1$ at the end of this section.

\begin{figure}[tb]
\begin{center}
\includegraphics[width=160mm]{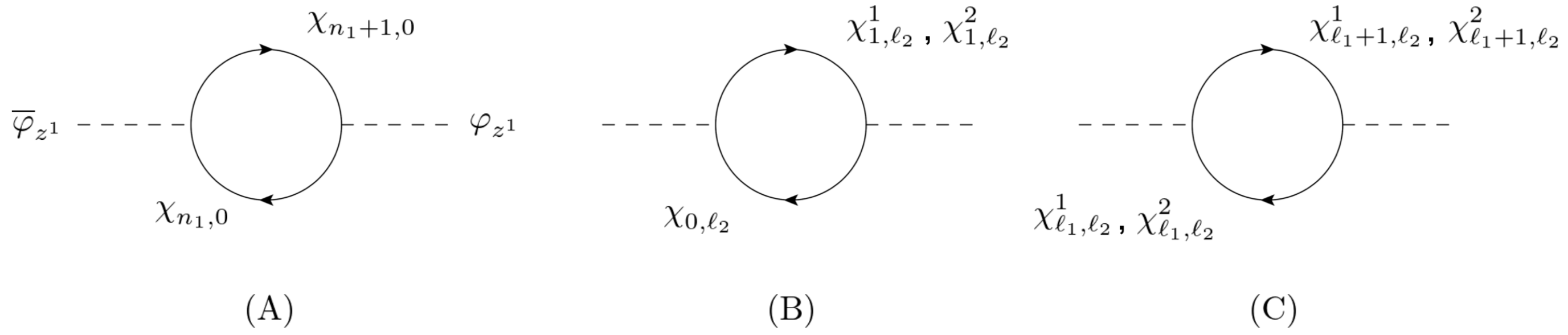}
\caption{one-loop diagrams for  
quantum corrections to the squared mass of $\varphi_{z^1}$.
(A), (B), and (C) correspond to the first, second, and third terms in the right-hand side of Eq. (5.1).}
\label{fig:1loopMass_phi1}
\end{center}
\end{figure}

\begin{figure}[tb]
\begin{center}
\includegraphics[width=160mm]{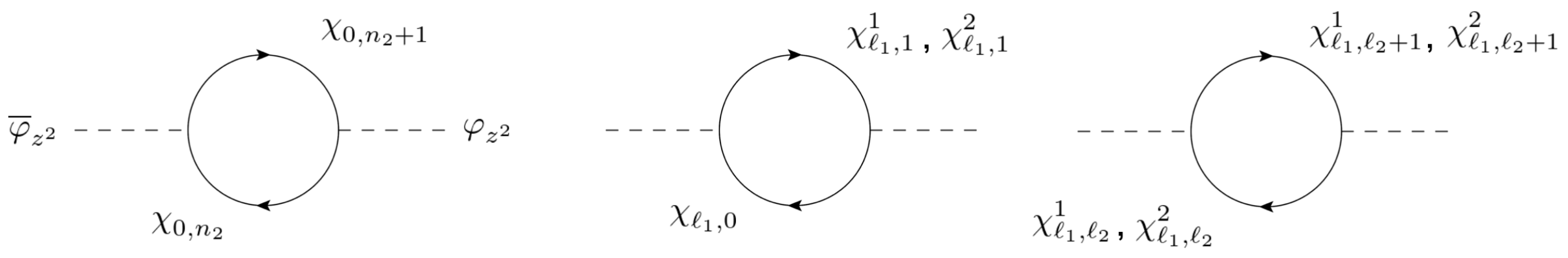}
\caption{one-loop diagrams for  
quantum corrections to the squared mass of $\varphi_{z^2}$.}
\label{fig:1loopMass_phi2}
\end{center}
\end{figure}

We calculate the quantum correction to the squared mass of $\varphi_{z^1}$ at the one-loop level, 
denoted by $\delta m_{\varphi_{z^1}}^2$. 
The Feynman diagrams that contribute to the quantum correction are listed 
in Fig.~\ref{fig:1loopMass_phi1}.
%
Extracting relevant interaction vertices from Eq.~\eqref{eq:r1r2_intPsiYukawa-6}, 
$\delta m_{\varphi_{z^1}}^2$ is calculated as 
\begin{align}
\label{eq:loop-correction_phi1-2}
\delta m_{\varphi_{z^1}}^2
=
&\,+
2(\sqrt{2}q_1)^2  
\sum_{n_1=0}^\infty
\int \frac{{\rm d}^4k}{(2\pi)^4} 
\frac{k^2}{\big( k^2 + M_{n_1,0}^2 \big)\big( k^2 + M_{n_1+1,0}^2 \big)}
\nn\\
&\,+
4(\sqrt{2}q_1)^2  
\sum_{\ell_2=1}^\infty
\int \frac{{\rm d}^4k}{(2\pi)^4} 
\frac{
k^2 + s_{2\theta_{0,\ell_2}}^{} s_{2\theta_{1,\ell_2}}^{} 
M_{0,\ell_2} M_{1,\ell_2}}{\big(k^2 + M_{0,\ell_2}^2\big) \big(k^2 + M_{1,\ell_2}^2\big)}
\nn\\
&\,+
4(\sqrt{2}q_1)^2
\sum_{\ell_1=1}^\infty
\sum_{\ell_2=1}^\infty
\int \frac{{\rm d}^4k}{(2\pi)^4} 
\frac{k^2
+ s_{2\theta_{\ell_1,\ell_2}}^{} s_{2\theta_{\ell_1+1,\ell_2}}^{} M_{\ell_1,\ell_2} M_{\ell_1+1,\ell_2}}{
\big( k^2 + M_{\ell_1,\ell_2}^2 \big)\big( k^2 + M_{\ell_1+1,\ell_2}^2 \big)} \, ,
\end{align} 
where $q_i=q \rho_i/\sqrt{\mathrm{Vol}(T^4)}$.
The charge $q_i$ is dimensionless in 4D since $q$ has a mass dimension $-3$.
The first, second, and third terms in the right-hand side of 
Eq.~\eqref{eq:loop-correction_phi1-2} correspond to the one-loop amplitudes 
for Feynman diagrams (A), (B), and (C) in Fig.~\ref{fig:1loopMass_phi1}, respectively. 
From Eq.~\eqref{eq:4D_tan2theta}, we also have the following relations,
\begin{align}
\label{eq:loop-correction_sin-cos-relation}
s_{2\theta_{\ell_1,\ell_2}}
= \sin2\theta_{\ell_1,\ell_2}
= \frac{M_{0,\ell_2}}{M_{\ell_1,\ell_2}} \, .
\end{align}
Replacing $s_{2\theta_{\ell_1,\ell_2}}$ by mass ratios 
in Eq.~\eqref{eq:loop-correction_phi1-2}, we obtain 
\begin{align}
\label{eq:loop-correction_phi1}
\delta m_{\varphi_{z^1}}^2
=&\,
+2(\sqrt{2}q_1)^2 
\int \frac{{\rm d}^4k}{(2\pi)^4} 
\sum_{n_1=0}^\infty
\frac{K^2}{\big(K^2 + M_{n_1,0}^2\big)\big(K^2 + M_{n_1+1,0}^2 \big)}
\Bigg|_{K^2=k^2}
\nn\\
&\,+
4(\sqrt{2}q_1)^2
\sum_{n_2=0}^\infty
\int \frac{{\rm d}^4k}{(2\pi)^4} 
\sum_{n_1=0}^\infty
\frac{K^2}{\big(K^2 + M_{n_1,0}^2 \big)\big(K^2 + M_{n_1+1,0}^2 \big)}
\Bigg|_{K^2=k^2+M_{0,n_2+1}^2},
\end{align}
where the second and third terms in Eq.~\eqref{eq:loop-correction_phi1-2} 
are combined appropriately. 
To evaluate these quantum corrections, we decompose the integrand into partial fractions:
\begin{align}
\label{calculatepart2}
\sum_{n_1=0}^\infty\frac{K^2}{\big(K^2+M_{n_1,0}^2\big)\big(K^2+M_{n_1+1,0}^2\big)}
&=\sum_{n_1=0}^\infty\l(\frac{n_1 + 1}{\big(K^2+M_{n_1+1,0}^2 \big)}
-\frac{n_1}{\big(K^2+M^2_{n_1,0}\big)}\r) \nn \\
&=0\,.
\end{align}
Using the shift $n_1\rightarrow n_1+1$ in the second term, we find that 
quantum corrections to $\varphi_{z^1}$ cancel out completely at the one-loop level as
\begin{align}
\delta m^2_{\varphi_{z^1}}=0\,.
\end{align}
We also confirm the cancellation of the one-loop contribution to the squared mass of $\varphi_{z^2}$. 
The corresponding Feynman diagrams are shown in Fig.~\ref{fig:1loopMass_phi2}. 
The one-loop corrections are obtained by replacing $M_{n_1,0}$ with $M_{0,n_2}$, 
$n_1$ with $n_2$, and $q_1$ with $q_2$ in Eq.\eqref{eq:loop-correction_phi1}. 
Note that $s_{2\theta_{n_1,n_2}}$ is converted to $c_{2\theta_{n_2,n_2}}$ under this replacement. 
Following the above procedure, we find 
\begin{align}
\label{eq:1loopMass_phi2-cancel}
\delta m^2_{\varphi_{z^2}}=0 \, . 
\end{align}

We comment on the case of ${\rm rank}\VEV{F}=1$. 
In this case, the quantum corrections to the squared mass for $\varphi_{z^1}$ are given by 
\begin{align}
\label{rankF=1phiz1}
\delta m^2_{\varphi_{z^1}}&=4(\sqrt{2}q_1)^2\sum_{l,m=-\infty}^\infty
\int\frac{{\rm d}^4k}{(2\pi)^4}\, \sum_{n_1=0}^\infty 
\frac{K^2}{\big(K^2+M_{n_1,0}^2\big)\big(K^2+M_{n_1+1,0}^2\big)}
\Bigg|_{K^2=k^2+\frac{l^2+m^2}{R^2_2}}\,, 
\end{align}
where $l,m\in\mathbb{Z}$. 
Therefore, we obtain
\begin{align}
\delta m^2_{\varphi_{z^1}}=0\,.
\end{align}
On the other hand, that for $\varphi_{z^2}$ is calculated as
\begin{align}
\label{eq:4D_1loopMass-phi2:diverge}
\delta m^2_{\varphi_{z^2}}&=+2(\sqrt{2}q_2)^2\sum_{l,m=-\infty}^\infty
\int\frac{{\rm d}^4k}{(2\pi)^4}\frac{k^2}{\l(k^2+\frac{l^2+m^2}{R^2_2}\r)^2} \nn \\
&\quad+4(\sqrt{2}q_2)^2\sum_{\ell_1=1}^\infty\sum_{l,m=-\infty}^\infty
\int\frac{{\rm d}^4k}{(2\pi)^4}\frac{k^2+M_{\ell_1,0}^2}{\l(k^2+M_{\ell_1,0}^2+\frac{l^2+m^2}{R^2_2}\r)^2}\,.
\end{align}
These corrections do not cancel out and do diverge\footnote{An expression similar to 
the first term in Eq.\eqref{eq:4D_1loopMass-phi2:diverge} is found 
in $T^2$ without magnetic flux case. 
See \cite{Antoniadis:2001cv,Buchmuller:2016gib}.}.
This means that $\varphi_{z^2}$ cannot be a NG mode. 

\section{Summary and discussions}
\label{sec:summary}
%
We have studied a $U(1)$ gauge theory on $M^4\times T^{2n}$ with background magnetic fluxes. 
For $n \geq 2$, multiple magnetic fluxes can be introduced, including off-diagonal fluxes.
We have shown that even in the presence of off-diagonal fluxes, one can always take 
a basis with only diagonal fluxes by an appropriate coordinate transformation. 
It is found that the number of independent magnetic fluxes is determined by the rank of 
the classical value of the field strength matrix, ${\rm rank}\VEV{F}$. 
The number of NG modes (WL scalar fields) is also determined by ${\rm rank}\VEV{F}$, 
where WL scalar fields are zero modes induced from extra components of the higher-dimensional 
gauge field. 
We have constructed a Landau level of fermions by introducing creation and annihilation 
operators in the flux diagonal basis. 
As examples, we derived the mass spectrum of four-dimensional fermions in $T^4$ and 
$T^6$ extra-dimensions. 

In Ref.~\cite{Buchmuller:2018eog},  the cancellation of the quantum corrections to the mass 
of the WL scalar field were proven in a $U(1)$ gauge theory on $M^4\times T^2$. 
In this paper, we have explicitly shown that the quantum corrections to the mass of 
the WL scalar field cancel out in the case of $T^4$ with arbitrary magnetic fluxes. 
Using our developed procedure, this result is easily extended to the case with $T^{2n}$.

We comment on an issue that needs to be resolved when interpreting this theory as a 
realistic model for the Higgs boson in the SM. 
As a mechanism to give a mass to the WL scalar field, we will consider three possibilities. 
First, when we extend our analysis to toroidal orbifolds, the mass will be generated 
by the breaking of the shift symmetry. 
We will leave a comprehensive study about NG modes on toroidal orbifolds in the future. 
Second, these WL scalar fields would acquire their masses by the Hosotani mechanism \cite{Hosotani:1983xw}. 
However, it was pointed out in Ref. \cite{Honda:2019ema} that the effective potential is independent of the WL phase in a six-dimensional $U(1)$ theory on $T^2$ with magnetic flux. 
Thus, it would be interesting to analyze the effective potential in $T^{2n}$ and 
its orbifold backgrounds. 
Finally, from the viewpoint of string theory, the stabilization of such massless modes 
(open string moduli) would be realized in M-theory and F-theory compactifications 
with background four-form fluxes (see, e.g.,  \cite{Becker:1996gj,Sethi:1996es,Gukov:1999ya,Denef:2005mm,Honma:2017uzn}). 
However, an arbitrary value of background fluxes including magnetic fluxes discussed 
in this paper is not allowed by the cancellation of D-brane charges. 
Rather, it is bounded from above in string models \cite{Ishiguro:2020rot}. 
It would be interesting to realize the stabilization of WL scalar fields 
in string models, which is left for future work.

\section*{Acknowledgments}

This work was supported in part by JSPS Grant-in-Aid for Scientific Research KAKENHI Grants No.~JP22K03620, JP18H05543 (K.T.), JP20K14477 (H.O.), JP23H04512 (H.O.), JP22KJ2621 (T.H.), by National Natural Science Foundation of China Grant No.~NSFC-12347112 (Y.U.), and by the Guangdong Major Project of Basic and Applied Basic Research No.~2020B0301030008 (Y.U.).
\\

\appendix
\def\thesection{Appendix.\Alph{section}}
\setcounter{section}{0}
\section{\label{app-A}}
\renewcommand{\theequation}{{\rm{A}}.\arabic{equation}}
\setcounter{equation}{0}
We derive the Yukawa interactions among WL scalar fields and fermions.
The Yukawa interactions in four-dimensional theory are given by
\begin{align}
\label{eq:r1r2_intPsiYukawa-6}
&
\int_{T^4}
{\rm d}^4z \,\sqrt{{\rm det}h}\,
\overline{\Psi} i\sla{D}_z \Psi \bigg|_{\varphi}
\nn\\
=&\,
+\sqrt{2}q_1\sum_{n_1=0}^\infty
\varphi_{z^1} \overline{\chi}_{n_1+1,0}^{} P_L \chi_{n_1,0}^{}
\nn\\
&\,+
\sqrt{2}q_1
\sum_{\ell_2=1}^\infty
\varphi_{z^1} 
              \Big\{  \overline{\chi}^1_{1,\ell_2}
                        (c_{\theta_{1,\ell_2}} P_R - s_{\theta_{1,\ell_2}}P_L) \chi_{0,\ell_2}^{}
                        +
                        \overline{\chi}^2_{1,\ell_2}
                        (s_{\theta_{1,\ell_2}} P_R - c_{\theta_{1,\ell_2}} P_L) \chi_{0,\ell_2}^{}
              \Big\} 
\nn\\
&\,
+\sqrt{2}q_1
\sum_{\ell_1=1}^\infty
\sum_{\ell_2=1}^\infty
\varphi_{z^1} 
               \Big\{
                   \overline{\chi}^1_{\ell_1+1,\ell_2}
                       ( +c_{\theta_{\ell_1+1,\ell_2}} c_{\theta_{\ell_1,\ell_2}} P_R
                          -s_{\theta_{\ell_1+1,\ell_2}} s_{\theta_{\ell_1,\ell_2}} P_L  )
                       \chi_{\ell_1,\ell_2}^1  
                   \nn\\
                   &\qquad\qquad\qquad\qquad\qquad~
                   +\overline{\chi}^1_{\ell_1+1,\ell_2}
                       ( -c_{\theta_{\ell_1+1,\ell_2}}s_{\theta_{\ell_1,\ell_2}} P_R  
                        +s_{\theta_{\ell_1+1,\ell_2}}c_{\theta_{\ell_1,\ell_2}} P_L  )
                       \chi_{\ell_1,\ell_2}^2 
                   \nn\\
                   &\qquad\qquad\qquad\qquad\qquad~      
                   +\overline{\chi}^2_{\ell_1+1,\ell_2}
                       ( +s_{\theta_{\ell_1+1,\ell_2}}c_{\theta_{\ell_1,\ell_2}} P_R
                         -c_{\theta_{\ell_1+1,\ell_2}}s_{\theta_{\ell_1,\ell_2}} P_L  )
                       \chi_{\ell_1,\ell_2}^1
                   \nn\\
                   &\qquad\qquad\qquad\qquad\qquad~
                   +\overline{\chi}^2_{\ell_1+1,\ell_2}
                       ( -s_{\theta_{\ell_1+1,\ell_2}}s_{\theta_{\ell_1,\ell_2}} P_R
                         +c_{\theta_{\ell_1+1,\ell_2}}c_{\theta_{\ell_1,\ell_2}} P_L  )
                       \chi_{\ell_1,\ell_2}^2
               \Big\}
\nn\\
&\,
-\sqrt{2}q_2
\varphi_{z^2} \overline{\chi}_{0,1} \chi_{0,0}^{}
+\sqrt{2}q_2
\sum_{\ell_2=1}^\infty
\varphi_{z^2} \overline{\chi}_{0,\ell_2+1}^{} P_L \chi_{0,\ell_2}^{}
\nn\\
&\,
+\sqrt{2}q_2
\sum_{\ell_1=1}^\infty
\varphi_{z^2} 
             \Big\{    
                      \overline{\chi}^1_{\ell_1,1} 
                      (c_{\theta_{\ell_1,1}} P_R -c_{\theta_{\ell_1,1}} P_L) \chi_{\ell_1,0}^2
                      + 
                      \overline{\chi}^2_{\ell_1,1} 
                      (s_{\theta_{\ell_1,1}} P_R +s_{\theta_{\ell_1,1}}P_L) \chi_{\ell_1,0}^2
             \Big\}                      
\nn\\
&\,
+\sqrt{2}q_2
\sum_{\ell_1=1}^\infty
\sum_{\ell_2=1}^\infty
\varphi_{z^2} 
               \Big\{ \overline{\chi}^1_{\ell_1,\ell_2+1}
                        (+c_{\theta_{\ell_1,\ell_2+1}}s_{\theta_{\ell_1,\ell_2}} P_R
                         +c_{\theta_{\ell_1,\ell_2+1}}s_{\theta_{\ell_1,\ell_2}} P_L)                      
                        \chi_{\ell_1,\ell_2}^1 
                        \nn\\
                   &\qquad\qquad\qquad\qquad\qquad~
                      +\overline{\chi}^1_{\ell_1,\ell_2+1}
                        (+c_{\theta_{\ell_1,\ell_2+1}} c_{\theta_{\ell_1,\ell_2}} P_R
                         -c_{\theta_{\ell_1,\ell_2+1}} c_{\theta_{\ell_1,\ell_2}} P_L)\chi_{\ell_1,\ell_2}^2
                        \nn\\
                   &\qquad\qquad\qquad\qquad\qquad~
                      +\overline{\chi}^2_{\ell_1,\ell_2+1}
                        (+s_{\theta_{\ell_1,\ell_2+1}} s_{\theta_{\ell_1,\ell_2}} P_R 
                         -s_{\theta_{\ell_1,\ell_2+1}} s_{\theta_{\ell_1,\ell_2}} P_L) \chi_{\ell_1,\ell_2}^1
                        \nn\\
                   &\qquad\qquad\qquad\qquad\qquad~
                      +\overline{\chi}^2_{\ell_1,\ell_2+1}
                        (+s_{\theta_{\ell_1,\ell_2+1}} c_{\theta_{\ell_1,\ell_2}} P_R 
                         +s_{\theta_{\ell_1,\ell_2+1}} c_{\theta_{\ell_1,\ell_2}} P_L) \chi_{\ell_1,\ell_2}^2    
               \Big\}
\nn\\
&\,
+\mathrm{H.c.}\,,
\end{align}
where $q_i=q \rho_i/\sqrt{\mathrm{Vol}(T^4)}$.

\def\thesection{Appendix.\Alph{section}}
\section{\label{app-B}}
\renewcommand{\theequation}{{\rm{B}}.\arabic{equation}}
\setcounter{equation}{0}

In this appendix, we derive the fermion mass spectrum on $M^4 \times T^{6}$ 
with background magnetic fluxes as a demonstration. 
The extension of the method to $T^{2n}$ is straightforward.  
Let us introduce dimensionless complex coordinates 
$z^i,~\bar{z}^{\,\bar{i}}$ $(\bar{i}=1,2,3)$ as
\begin{align}
\label{eq:flux_xyz-relation:appendixB}
z^i 
= x^i + \Omega^i{}_j\, y^j \, ,
\quad
\bar{z}^{\,\bar{i}} 
= x^{\,\bar{i}} 
+ \overline{\Omega}^{\,\bar{i}}{}_{\bar{j}} \, y^{\,\bar{j}} \, ,
\end{align}
where $x^i$, $y^i$ are real and dimensionless coordinates, 
and $\Omega^i{}_{j}(\in \mathbb{C}^3)$ denotes the complex structure. 
A free lagrangian for the Weyl fermion 
$\Psi$ on $M^4\times T^6$
with background magnetic flux is given by 
\begin{align}
\label{eq:t6_lag10D}
\mathcal{L}^{\rm 10D}_{\rm free}
=&\,
\overline{\Psi} i\Gamma^\mu \partial_\mu \Psi
+
\overline{\Psi} 
i \sla{\mathcal{D}}_z^{T^6} \Psi \, ,
\end{align}
with
\begin{align}
i\sla{\mathcal{D}}_z^{T^6}
&=
i \sum_{j=1}^3\Gamma^{z^j} \mathcal{D}_{z^j} 
+ i \sum_{\bar{j}=1}^3 \Gamma^{\bar{z}^{\bar{j}}} \mathcal{D}_{\bar{z}^{\bar{j}}}  
\nn\\
&=
2\,i 
\l(  \begin{array}{cccccccc}
     0 & \rho_3\mathcal{D}_{z^3} & \rho_2\mathcal{D}_{z^2} & 0 & \rho_1\mathcal{D}_{z^1} & 0 & 0 & 0
     \\
     \rho_3\mathcal{D}_{\bar{z}^3} & 0 & 0 & -\rho_2\mathcal{D}_{z^2} & 0 & -\rho_1\mathcal{D}_{z^1} & 0 & 0
     \\
     \rho_2\mathcal{D}_{\bar{z}^2} & 0 & 0 & \rho_3\mathcal{D}_{z^3} & 0 & 0 & -\rho_1\mathcal{D}_{z^1} & 0
     \\
     0 & -\rho_2\mathcal{D}_{\bar{z}^2} & \rho_3D_{\bar{z}^3} & 0 & 0 & 0 & 0 & \rho_1D_{z^1}
     \\
     \rho_1\mathcal{D}_{\bar{z}^1} & 0 & 0 & 0 & 0 & \rho_3\mathcal{D}_{z^3} & \rho_2\mathcal{D}_{z^2} & 0
     \\
     0 & -\rho_1\mathcal{D}_{\bar{z}^1} & 0 & 0 & \rho_3\mathcal{D}_{\bar{z}^3} & 0 & 0 & -\rho_2\mathcal{D}_{z^2}
     \\
     0 & 0 & -\rho_1\mathcal{D}_{\bar{z}^1} & 0 & \rho_2\mathcal{D}_{\bar{z}^2} & 0 & 0 & \rho_3\mathcal{D}_{z^3}
     \\
     0 & 0 & 0 & \rho_1\mathcal{D}_{\bar{z}^1} & 0 & -\rho_2\mathcal{D}_{\bar{z}^2} & \rho_3\mathcal{D}_{\bar{z}^3} & 0
     \end{array}
\r) \, ,      
\end{align}
where $\rho_i = 1/(2\pi R_i)$ for all $i=1,2,3$ and $R_i$ are scale factors in $T^6$. 
We use the following Gamma matrices on the complex coordinates of $T^6$:
\begin{align}
&\Gamma^{z^1}
= \sigma_+ \otimes \sigma^3 \otimes \sigma^3\, ,
\qquad
\Gamma^{z^2}
= I \otimes \sigma_+ \otimes \sigma^3\,,
\qquad
\Gamma^{z^3}
= I \otimes I \otimes \sigma_+\,,
\nn\\
&\Gamma^{\bar{z}^1}
= \sigma_- \otimes \sigma^3 \otimes \sigma^3\, ,
\qquad
\Gamma^{\bar{z}^2}
= I \otimes \sigma_- \otimes \sigma^3\, , 
\qquad
\Gamma^{\bar{z}^3}
= I \otimes I \otimes \sigma_-\,
.
\end{align}
Note that $\Psi$ is expressed in terms of four-component fields as 
\begin{align}
\label{eq:DiracEq_Psi10D}
\Psi
=
\l(  \begin{array}{c}
     \psi^{\bf 1}_L
     \\
     \psi^{\bf 2}_R
     \\
     \psi^{\bf 3}_R
     \\
     \psi^{\bf 4}_L
     \\
     \psi^{\bf 5}_R
     \\
     \psi^{\bf 6}_L
     \\
     \psi^{\bf 7}_L
     \\
     \psi^{\bf 8}_R
     \end{array}
\r) \, ,
\end{align}
where $L$ and $R$, respectively, correspond to the left- and right-handed chiralities 
in 4D after integrating the extra dimension $T^6$. 
The chirality matrix $\Gamma^5$ is given by $\Gamma^5=\sigma^3\otimes \sigma^3 \otimes \sigma^3$.

Similar to Sec.~\ref{sec:DiracEq}, the commutation relations for covariant derivatives are given below,
\begin{align}
\label{eq:diagonal_DDcomm-t6}
[\mathcal{D}_{z^i}, \mathcal{D}_{\bar{z}^{\bar{j}}}] 
=
-q \VEV{F_{z^i\bar{z}^{\bar{j}}}} \, .
\end{align}
In the following, we take the flux diagonal basis. 
Based on the above commutation relations, we define creation and annihilation operators 
by normalizing the covariant derivatives,
\begin{align}
\label{eq:diagonal_a-def-t6}
&a_1^{} 
= \frac{i}{\sqrt{q \VEV{F_{z^1\bar{z}^1}}}}\mathcal{D}_{z^1} \, ,
\quad~~
a_1^\dagger 
= \frac{i}{\sqrt{q \VEV{F_{z^1\bar{z}^1}}}}\mathcal{D}_{\bar{z}^1} \, ,
\nn\\
&a_2^{} 
= \frac{i}{\sqrt{q \VEV{F_{z^2\bar{z}^2}}}}\mathcal{D}_{z^2} \, ,
\quad~~
a_2^\dagger 
= \frac{i}{\sqrt{q \VEV{F_{z^2\bar{z}^2}}}}\mathcal{D}_{\bar{z}^2} \, ,
\nn\\
&a_3^{} 
= \frac{i}{\sqrt{q \VEV{F_{z^3\bar{z}^3}}}}\mathcal{D}_{z^3} \, ,
\quad~~
a_3^\dagger 
= \frac{i}{\sqrt{q \VEV{F_{z^3\bar{z}^3}}}}\mathcal{D}_{\bar{z}^3} \, , 
\end{align}
where $q\VEV{F_{z^1\bar{z}^1}}>0$, $q\VEV{F_{z^2\bar{z}^2}}>0$ and $q\VEV{F_{z^3\bar{z}^3}}>0$ are assumed. 
The operators satisfy $[a_i^{},a_j^\dagger]=\delta_{ij}$, and
$[a_i^{},a_j^{}]=[a_i^\dagger,a_j^\dagger]=0$. 
The ground state mode function $\xi_{0,0,0}$ is defined by $a_i\,\xi_{0,0,0}=0$. 
Acting $\xi_{0,0,0}$ on the creation operators $a^\dag_i$, the mode function $\xi_{n_1,n_2,n_3}$ is obtained as
\begin{align}
    \xi_{n_1,n_2,n_3}=\frac{1}{\sqrt{n_1! n_2!n_3!}}(a_1^\dagger)^{n_1}(a_2^\dagger)^{n_2}(a_3^\dagger)^{n_3}\xi_{0,0,0}\,.
\end{align}
This mode function $\xi_{n_1,n_2,n_3}$ is an eigenstate for the number operators
$a_1^\dagger a_1^{}$, $a_2^\dagger a_2^{}$, and $a_3^\dagger a_3^{}$,
\begin{align}
\label{eq:diraceqkk_nxi-t6}
a_i^\dagger a_i^{} \xi_{n_1,n_2} = n_i \xi_{n_1,n_2,n_3} 
\quad
(i=1,2,3) \, ,
\end{align}
where repeated indices on the left side are not summed.
The mode function $\xi_{n_1,n_2,n_3}$ is normalized and satisfies an orthogonality condition,
\begin{align}
\label{eq:DiracEq_orthogonal10D}
\int_{T^6}
{\rm d}^6z\,\sqrt{{\rm det}h}\,
\overline{\xi}_{n_1,n_2,n_3}\xi_{n'_1,n'_2,n'_3}
=
\delta_{n_1n'_1}
\delta_{n_2n'_2}
\delta_{n_3n'_3} \, .
\end{align}
The mode function satisfies the following relations:
\begin{align}
\label{eq:diraceqkk_n+1_n-t6}
&a_1 \xi_{n_1,n_2,n_3} = \sqrt{n_1}\, \xi_{n_1-1,n_2,n_3} \, ,
\qquad
a_1^\dagger \xi_{n_1,n_2,n_3} = \sqrt{n_1+1}\, \xi_{n_1+1,n_2,n_3} \, ,
\nn\\
&a_2 \xi_{n_1,n_2,n_3} = \sqrt{n_2}\, \xi_{n_1,n_2-1,n_3} \, ,
\qquad
a_2^\dagger \xi_{n_1,n_2,n_3} = \sqrt{n_2+1}\, \xi_{n_1,n_2+1,n_3} \, ,
\nn\\
&a_3 \xi_{n_1,n_2,n_3} = \sqrt{n_3}\, \xi_{n_1,n_2,n_3-1} \, ,
\qquad
a_3^\dagger \xi_{n_1,n_2,n_3} = \sqrt{n_3+1}\, \xi_{n_1,n_2,n_3+1} \, .
\end{align}
The squared Dirac operator becomes 
\begin{align}
(i\sla{\mathcal{D}}_z^{T^6})^2
=&\, 4\,q\, 
{\rm diag}
\big( \rho_1^2 \VEV{F_{z^1\bar{z}^1}} (a_1^\dag a_1+1)+\rho_2^2\VEV{F_{z^2\bar{z}^2}} (a_2^\dag a_2+1)
+\rho_3^2\VEV{F_{z^3\bar{z}^3}} (a_3^\dag a_3+1)\,, \nn \\
        &\qquad \qquad
\rho_1^2 \VEV{F_{z^1\bar{z}^1}} (a_1^\dag a_1+1)+\rho_2^2\VEV{F_{z^2\bar{z}^2}} (a_2^\dag a_2+1)
+\rho_3^2\VEV{F_{z^3\bar{z}^3}} a_3^\dag a_3\,, \nn \\
        &\qquad \qquad
\rho_1^2 \VEV{F_{z^1\bar{z}^1}} (a_1^\dag a_1+1)+\rho_2^2\VEV{F_{z^2\bar{z}^2}} a_2^\dag a_2
+\rho_3^2\VEV{F_{z^3\bar{z}^3}} (a_3^\dag a_3+1)\,, \nn \\
        &\qquad \qquad
\rho_1^2 \VEV{F_{z^1\bar{z}^1}} (a_1^\dag a_1+1)+\rho_2^2\VEV{F_{z^2\bar{z}^2}} a_2^\dag a_2
+\rho_3^2\VEV{F_{z^3\bar{z}^3}} a_3^\dag a_3\,, \nn \\
        &\qquad \qquad
\rho_1^2 \VEV{F_{z^1\bar{z}^1}} a_1^\dag a_1+\rho_2^2\VEV{F_{z^2\bar{z}^2}} (a_2^\dag a_2+1)
+\rho_3^2\VEV{F_{z^3\bar{z}^3}} (a_3^\dag a_3+1)\,, \nn \\
        &\qquad \qquad
\rho_1^2 \VEV{F_{z^1\bar{z}^1}} a_1^\dag a_1+\rho_2^2\VEV{F_{z^2\bar{z}^2}} (a_2^\dag a_2+1)
+\rho_3^2\VEV{F_{z^3\bar{z}^3}} a_3^\dag a_3\,, \nn \\
        &\qquad \qquad
\rho_1^2 \VEV{F_{z^1\bar{z}^1}} a_1^\dag a_1+\rho_2^2\VEV{F_{z^2\bar{z}^2}} a_2^\dag a_2
+\rho_3^2\VEV{F_{z^3\bar{z}^3}} (a_3^\dag a_3+1)\,, \nn \\
        &\qquad \qquad
\rho_1^2 \VEV{F_{z^1\bar{z}^1}} a_1^\dag a_1+\rho_2^2\VEV{F_{z^2\bar{z}^2}} a_2^\dag a_2
+\rho_3^2\VEV{F_{z^3\bar{z}^3}} a_3^\dag a_3
\big)\,. 
\end{align}
From the above, we see that only right-handed fermion 
$\psi^{\bf 8}_{R,n_1,n_2,n_3}$ has a massless mode $\psi^{\bf 8}_{R,0,0,0}$.
Note that ten-dimensional fermions are expanded with the mode function $\xi_{n_1,n_2,n_3}$ as
\begin{align}
\label{eq:diraceqkk_kkexpansion-t6}
\psi^{\bf 1}_L
&=
\sum_{n_1=0}^\infty
\sum_{n_2=0}^\infty
\sum_{n_3=0}^\infty
\psi^{\bf 1}_{L,n_1+1,n_2+1,n_3+1}
\xi_{n_1,n_2,n_3} \, ,
\nn\\
\psi^{\bf 2}_R
&=
\sum_{n_1=0}^\infty
\sum_{n_2=0}^\infty
\sum_{n_3=0}^\infty
\psi^{\bf 2}_{R,n_1+1,n_2+1,n_3}
\xi_{n_1,n_2,n_3} \, ,
\nn \\
\psi^{\bf 3}_R
&=
\sum_{n_1=0}^\infty
\sum_{n_2=0}^\infty
\sum_{n_3=0}^\infty
\psi^{\bf 3}_{R,n_1+1,n_2,n_3+1}
\xi_{n_1,n_2,n_3} \, ,
\nn \\
\psi^{\bf 4}_L
&=
\sum_{n_1=0}^\infty
\sum_{n_2=0}^\infty
\sum_{n_3=0}^\infty
\psi^{\bf 4}_{L,n_1+1,n_2,n_3}
\xi_{n_1,n_2,n_3} \, ,
\nn \\
\psi^{\bf 5}_R
&=
\sum_{n_1=0}^\infty
\sum_{n_2=0}^\infty
\sum_{n_3=0}^\infty
\psi^{\bf 5}_{R,n_1,n_2+1,n_3+1}
\xi_{n_1,n_2,n_3} \, ,
\nn\\
\psi^{\bf 6}_L
&=
\sum_{n_1=0}^\infty
\sum_{n_2=0}^\infty
\sum_{n_3=0}^\infty
\psi^{\bf 6}_{L,n_1,n_2+1,n_3}
\xi_{n_1,n_2,n_3} \, ,
\nn \\
\psi^{\bf 7}_L
&=
\sum_{n_1=0}^\infty
\sum_{n_2=0}^\infty
\sum_{n_3=0}^\infty
\psi^{\bf 7}_{L,n_1,n_2,n_3+1}
\xi_{n_1,n_2,n_3} \, ,
\nn \\
\psi^{\bf 8}_R
&=
\sum_{n_1=0}^\infty
\sum_{n_2=0}^\infty
\sum_{n_3=0}^\infty
\psi^{\bf 8}_{R,n_1,n_2,n_3}
\xi_{n_1,n_2,n_3} \, , 
\end{align}
where the 4D fermions $\psi_{L/R, n_1,n_2,n_3}^{\bf 1,2,3,4,5,6,7,8}$ are 
defined to have masses of 
\begin{align}
M_{n_1,n_2,n_3}=2 \sqrt{q\Big(
\VEV{F_{z^1\bar{z}^1}} n_1+
\VEV{F_{z^2\bar{z}^2}} n_2+
\VEV{F_{z^3\bar{z}^3}} n_3 \Big)} \, .
\end{align}
%


\bibliographystyle{utphys28mod}
\bibliography{ref.bib}

\end{document}